\documentclass[prl,floatfix,showpacs,superscriptaddress,longbibliography,twocolumn]{revtex4-2}
\usepackage[USenglish]{babel}
\usepackage{amsmath,amssymb,amsfonts}
\usepackage{epstopdf}
\usepackage{epsfig} 
\usepackage{graphicx}
\usepackage[T1]{fontenc}
\usepackage{color}
\usepackage{url}
\usepackage{changes}

\usepackage[colorlinks=true,linkcolor=blue,citecolor=red]{hyperref}

\newcommand{\equ}[1]
{Eq.~(\ref{#1})}

\newcommand{\figu}[1]
{Fig.~\ref{#1}}

\makeatletter
\DeclareFontFamily{OMX}{MnSymbolE}{}
\DeclareSymbolFont{MnLargeSymbols}{OMX}{MnSymbolE}{m}{n}
\SetSymbolFont{MnLargeSymbols}{bold}{OMX}{MnSymbolE}{b}{n}
\DeclareFontShape{OMX}{MnSymbolE}{m}{n}{
    <-6>  MnSymbolE5
   <6-7>  MnSymbolE6
   <7-8>  MnSymbolE7
   <8-9>  MnSymbolE8
   <9-10> MnSymbolE9
  <10-12> MnSymbolE10
  <12->   MnSymbolE12
}{}
\DeclareFontShape{OMX}{MnSymbolE}{b}{n}{
    <-6>  MnSymbolE-Bold5
   <6-7>  MnSymbolE-Bold6
   <7-8>  MnSymbolE-Bold7
   <8-9>  MnSymbolE-Bold8
   <9-10> MnSymbolE-Bold9
  <10-12> MnSymbolE-Bold10
  <12->   MnSymbolE-Bold12
}{}

\let\llangle\@undefined
\let\rrangle\@undefined
\DeclareMathDelimiter{\llangle}{\mathopen}%
                     {MnLargeSymbols}{'164}{MnLargeSymbols}{'164}
\DeclareMathDelimiter{\rrangle}{\mathclose}%
                     {MnLargeSymbols}{'171}{MnLargeSymbols}{'171}
                     \makeatother

\def\a{\alpha}              
          
        \def\m{\mu}      \def\n{\nu}
                    \def\s{\sigma}
\def\t{\tau}           

\def\X{\mathcal{X}}

\def\PP{{\cal P}} 
\def\FF{{\cal F}}
\def\TT{{\cal T}}\def\NN{{\cal N}} 
 
\def\DD{{\cal D}}

\def\GG{{\cal G}}

\def\=={\equiv}

\def\qed{\raise1pt\hbox{\vrule height5pt width5pt depth0pt}}

\def\cG0{{\cal G}_0} 
\def\cG{{\cal G}}

 \def\qa{{\bf q}} 

\def\ia{{\bf i}}

\def\ka{{\bf k} \alpha}

 \def\Re{{\rm Re}}

\def\ns{\scriptscriptstyle N}
\def\ts{\scriptscriptstyle T}
\def\ss{\scriptscriptstyle S}
\def\rs{\scriptscriptstyle R}
\def\mf{\scriptscriptstyle MF}

\renewcommand{\vec}{\mathbf}

\newcommand{\feval}[1]{\left\llangle{#1}\right\rrangle}

\def\vD{\mathbf{\Delta}}

\def\ka{\mathbf{k}}
\def\qa{\mathbf{q}}

\begin{document}
\author{F. Paoletti} 
\affiliation{International School for Advanced Studies (SISSA), via Bonomea 265, 34136 Trieste, Italy}

\author{L. Fanfarillo}   
\affiliation{CNR-ISC, Via dei Taurini, Rome, Italy}
\affiliation{International School for Advanced Studies (SISSA), via Bonomea 265, 34136 Trieste, Italy}

\author{M. Capone}
\affiliation{International School for Advanced Studies (SISSA), via Bonomea 265, 34136 Trieste, Italy}
\affiliation{CNR-IOM, Istituto Officina dei Materiali,
Consiglio Nazionale delle Ricerche, Via Bonomea 265, 34136 Trieste, Italy}

\author{A. Amaricci}
\affiliation{CNR-IOM, Istituto Officina dei Materiali,
Consiglio Nazionale delle Ricerche, Via Bonomea 265, 34136 Trieste, Italy}

\title{Topological Gap Opening without Symmetry Breaking from Dynamical Quantum Correlations}

\begin{abstract}
  Topological phase transitions are typically associated with 
  the formation of gapless states. Spontaneous symmetry breaking can
  lead to a gap opening thereby obliterating the topological nature of
  the  system. Here we highlight a completely different destiny for a
  topological transition in presence of interaction.  Solving a
  Bernevig-Hughes-Zhang model with local interaction, we show that
  dynamical quantum fluctuations  can lead to the opening of a gap
  without any symmetry breaking.
  As we vary the interaction and the bare mass of the model,
  the continuous gapless topological transition turns into a
  first-order one, associated with the presence of massive Dirac
  fermion at the transition point showing a Gross-Neveu critical
  behaviour near the quantum critical endpoint.
  We identify the gap opening as a condensed matter analog of the
    Coleman-Weinberg mechanism of mass generation.
\end{abstract}

\maketitle
The discovery of symmetry protected topological phases of
matter~\cite{Bernevig2006S,Qi2010PT,Moore2010N,Hasan2010RMP,Bernevig2013,Wen2017RMP,Bradlyn2017N,Rachel2018ROPIP}
has enriched the landscape of phase transitions beyond the
conventional Landau paradigm of the
symmetry-breaking~\cite{Nambu2009RMP,Zhang2009PRB}.
In presence of a given  symmetry the 
possible electronic band structures of an insulator can be divided into
distinct equivalence classes, which can only be connected through the
continuous closure of the energy gap from both sides of the transition through 
a topological quantum phase transition (TQPT).
The corresponding formation of symmetry-protected massless Dirac
fermions at the transition is  a distinctive feature of the
topological insulators~\cite{Zhou2006NP,Hsieh-D.2008N,Wehling2014AP,Young2015PRL,Zahid-Hasan2015}.

The presence of interactions can change the above scenario and a
gapped state can appear also at the transition.  
The standard phenomenology for this to occur requires a spontaneous symmetry breaking (SSB)~\cite{Hasan2010RMP,Qi2011RMP,Ezawa2013SR,Rachel2016JOPCM,Jia2022NP,Amaricci2023PRB}. Indeed, breaking a continuous symmetry opens a gap in the energy spectrum or, equivalently, gives a finite mass for the Dirac electrons which is understood in terms of the Anderson-Higgs mechanism. 
Clearly, a SSB can lead to break any of the symmetry protecting the topological state, thus leaving behind a topologically trivial long-range ordered phase. A similar scenario can be described within a static mean-field picture in the channel where
SSB takes place.

A great deal of attention has been recently drawn in different fields~\cite{You2018PRX,Slagle2015PRB,Wang2022S} to 
novel possible mechanisms of spontaneous mass generation which preserve
the symmetry, beyond the conventional SSB description.
Here we show that such  process describes 
the gap opening for Dirac electrons at the boundary of a topological insulator. 
More concretely, we address the question whether or not
electron-electron interactions can drive the formation of a spontaneous mass for
the otherwise gapless electrons at a topological transition. The lack of a gap 
closing at the TQPT is expected to change
the character of the transition which becomes necessarily discontinuous despite the symmetries 
protecting the topological phase are preserved.

For the sake of definiteness we consider a two-dimensional Bernevig-Hughes-Zhang (BHZ) model augmented via the inclusion of local electron-electron interactions 
that preserve some 
symmetries of the model.  
Without interactions,
this model features a TQPT through the formation of a gapless state. The
control parameter of the transition is the energy splitting between
two electronic orbitals playing the role of a mass term. As a
consequence the difference in the occupation of the orbitals, or
orbital polarization, is expected to assume a prominent role.
Since the orbital symmetry is broken by the mass term, the concept of SSB does not apply to the TQPT. Within a mean-field theory, the interactions simply dress the mass term, shifting the topological transition without changing its nature with respect to the non-interacting limit. 

In this work we go beyond mean-field using a variational approach including quantum fluctuations not only of the  orbital polarization, but also in the other particle-hole channels. We demonstrate a new scenario in which a gap opens at the TQPT without breaking  any of the symmetries of the model. 
We show explicitly that the quantum fluctuations in the different channels make the TQPT discontinuous for sufficiently large interactions~\cite{Amaricci2015PRL,Yamaji2007JMMM}. The first-order line ends in a critical endpoint, where we show a Gross-Neveu quantum critical behavior as a function of the relevant coupling strength~\cite{Liu2021PRB}. As we shall discuss in the following, the mechanism we revealed is reminiscent of the Coleman-Weinberg (CW) theory of mass generation~\cite{Coleman1973PRD,Rau2018PRL}. 

We solve an interacting BHZ model on a square lattice~\cite{Bernevig2006S,Wu2006PRL,Qi2010PT,Qian2014S,Amaricci2015PRL}
\begin{equation}
\mathcal{H} = \sum_{\ka} \psi^\dagger_{\ka} H^0_\ka  \psi_{\ka} + \sum_\ia \mathcal{H}^\mathrm{int}_\ia
\label{int_BHZ}
\end{equation}
where  $\psi_{\ka} = [c_{ \ka 1 \uparrow}, c_{\ka 2 \uparrow},
c_{\ka  1\downarrow}, c_{ \ka 2 \downarrow}]^T$ and the operators $c_{ \ka \a \sigma}$
annihilate an electron with momentum $\ka$, orbital $\alpha=1, 2$
and  spin $\sigma=\uparrow,\downarrow$. 
If we define 
$\Gamma_{\m\n}= \sigma_\m\otimes
\tau_\n$ with   $\s_\m$ and $\t_\n$  the Pauli
matrices, respectively, in the spin and orbital subspaces
the single-particle Hamiltonian reads:
$H^0_\ka = [M-2t(\cos(k_x)+\cos(k_y))]\Gamma_{03} +
\lambda\sin(k_x)\Gamma_{31}+\lambda\sin(k_y)\Gamma_{02}$, where
$M\geq0$ is the energy separation between the two orbitals which plays the role of the 
mass term, $t$ and $\lambda$ are the intra- and inter-orbital hopping amplitudes.
The model is invariant under time-reversal $\TT$ and inversion $\PP$
symmetries, $U(1)$ spin rotation around the $z$ axis~\cite{Blason2020PRB,Budich2014PRL,Mazza2020PRL,Amaricci2023PRB}.
In the following we set our energy unit so that $2t=1$ and 
focus on the regime of two electrons per site, i.e. half-filling.
The non-interacting model has a continuous topological transition between a 
QSHI for $M< 2$ and a trivial BI for $M> 2$ through the formation of a gapless Dirac state at $M=2$~\cite{Bernevig2006S}.

We assume a generic local interaction which preserves inversion $\PP$ and $U(1)$ spin symmetry around the $z$ axis
~\cite{Hohenadler2013JOPCM,Rachel2016JOPCM}: 
\begin{equation}
  \mathcal{H}^{int}_\ia = 
    -\frac{g_{\ns}}{2}\hat{N}_\ia^2 -\frac{g_{\ts}}{2}\hat{T}_{z\ia}^2 
    -\frac{g_{\ss}}{2}\hat{S}_{z\ia}^2 - \frac{g_{\rs}}{2}\hat{R}_{z\ia}^2 
\label{interaction}
\end{equation}
where  $\hat{N}_\ia\!=\!\tfrac{1}{2}\psi^+_\ia\Gamma_{00}\psi_\ia$ is half of
the total occupation per site, 
$\hat{T}_{z\ia}\!=\!\tfrac{1}{2}\psi^+_\ia\Gamma_{03}\psi_\ia$ and
$\hat{S}_{z\ia}\!=\!\tfrac{1}{2}\psi^+_\ia\Gamma_{30}\psi_\ia$ are,
respectively, the $z$ component of the orbital polarization and the spin operators and
$\hat{R}_{z\ia}=\tfrac{1}{2}\psi^+_\ia\Gamma_{33}\psi_\ia$; $\psi_\ia$ is the Fourier transform of $\psi_\ka$.
In the numerical calculations we will consider $g_{\ns} = -(3U-5J)$, $g_{\ts}=U-5J$, $g_{\ss} = U+J$ and $g_{\rs}  =U-J$ in order to recover the density-density version of the popular Kanamori-Hubbard~\cite{Georges2013ACMP} model used in a variety of works to study the interplay between the Hubbard $U$ and the Hund's exchange $J$ and their effect on TQPTs~\cite{Werner2007PRL,Budich2012PRB,Amaricci2015PRL}. 
We consider non-magnetic solutions in order to study $\mathcal{T}$ symmetry preserving transitions~\cite{Amaricci2016PRBa,Amaricci2017PRB}.
To simplify the notation in the following we define $\Gamma_{a=\ns,\ts,\ss,\rs}$ as the set $\frac{1}{2}\Gamma_{\m\n=00,03,30,33}$ in order to highlight the different channels.

The starting point of our analysis is to rewrite
the partition function of the interacting model (\ref{int_BHZ})
in terms of an effective problem coupled to space- and time-dependent (real) bosonic fields $\vD_q= \Delta_q^{a=\ns,\ts,\ss,\rs}$,
by performing an Hubbard–Stratonovich transformation.
The partition function reads: $\mathcal{Z} \equiv e^{-\beta \mathcal{F}} =  \int   
  \mathcal{D}\vD \ e^{-\beta \mathcal{N}  F[\vD]}$ in terms of the free energy
  functional 
\begin{equation}
  F[\vD] =
  \sum_{a q} \frac{|\Delta^a_q|^2 }{2 g_a} - \frac{1}{\beta
    \mathcal{N}} \text{Tr} \ln (-\GG^{-1}_{kq}) 
\label{Feff}
\end{equation}
where $q=(\qa,i\nu_m)$, $k=(\ka,i\omega_n)$, with $\qa, \ \ka$ wave
vectors in the first Brillouin zone and  $\nu_m, \ \omega_n$ 
the bosonic and fermionic Matsubara frequencies, respectively, 
$\beta$ is the inverse temperature and $\NN$ the total number of sites. 
$\text{Tr}$ indicates the trace over momentum, frequency, orbital and spin.
Finally $\GG_{kq} = \left(i\omega_n + \mu -
  H^0_\ka\delta_{\ka,\ka-\qa} -V_q \right)^{-1}$ is the interacting one-body Green's
function, where $V_q=-\sum_a\Delta^a_q \Gamma_a$ is an effective time-dependent potential depending on $\vD_q$.

\begin{figure*}
\includegraphics[width=\linewidth]{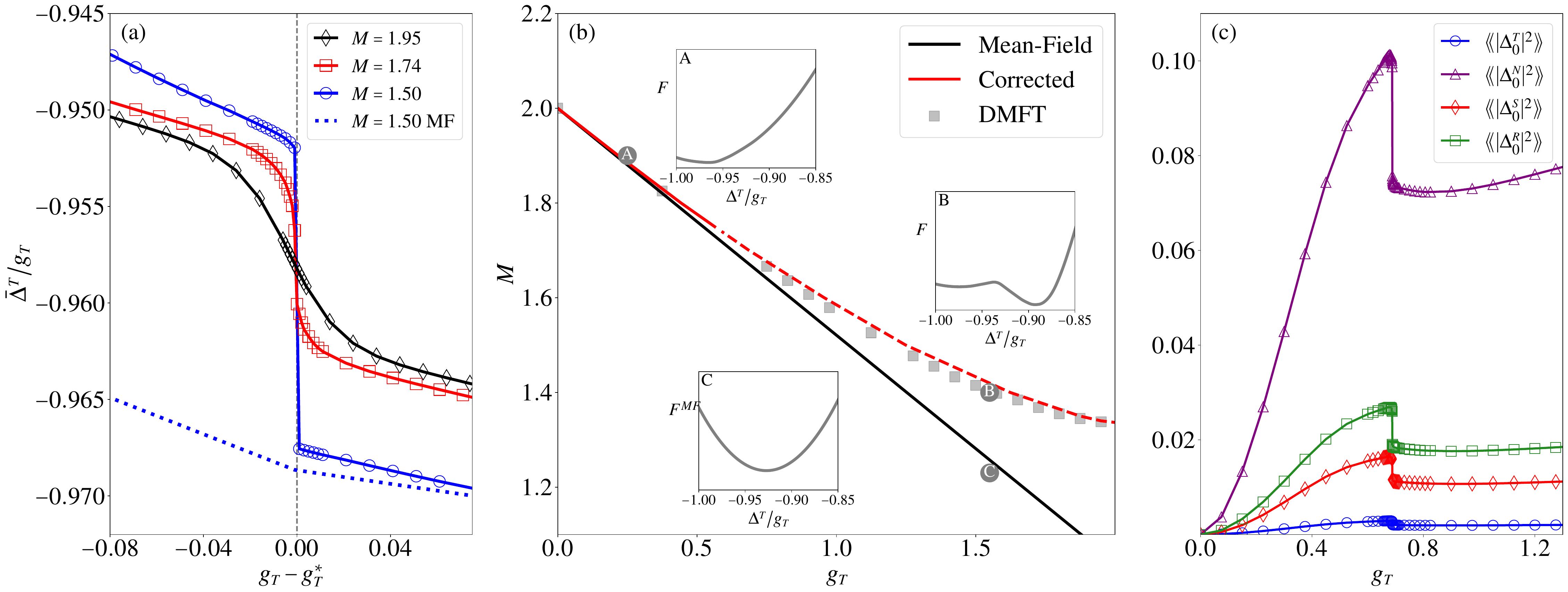}
\caption{(Color online)
  (a) Orbital polarization $\bar{\Delta}^{\ts}/g_{\ts}$ as a function of
 $g_{\ts}$ measured with respect to the TQPT point $g^*_{\ts}$.
  The open symbols correspond to the fluctuation-corrected results.
   The dotted line is the MF solution for $M=1.50$.  
   (b) Phase diagram in the $M$-$g_{\ts}$ plane comparing the
  topological transition line in the two approximations.
  The solid lines (black and red) denote a continuous TQPT, while the dashed
  line (red) marks a discontinuous one. Data from DMFT are indicated with filled symbols (gray).  
  The insets A, B and C show the free energy $F$ as a function of the
  orbital polarization for the three points marked on the curves. 
  (c) Static and homogeneous components $\feval{|\Delta^a_{q=0}|^2}$
  for $a=T,N,S,R$ of the potential fluctuations  as a function of
  $g_{\ts}$ across the topological transition for $M=1.70$.
}
\label{fig1}
\end{figure*}

The natural lowest-order approximation of \equ{Feff}
is a static mean-field (MF) solution, where the bosonic fields
$\Delta^a_q$ are approximated with time-independent and spatially
uniform quantities. The presence of the mass term breaks explicitly the symmetry between the orbitals, leading to a finite values of the orbital polarization $T_z$ already in the non-interacting model.
The non-magnetic solution ($\Delta^{\ss}_{\mf} = \Delta^{\rs}_{\mf} = 0$) at half-filling ($\Delta^{\ns}_{\mf}=1$), reduces to the single self-consistency equation $\Delta^{\ts} =
\frac{g_{\ts}}{\beta \mathcal{N}} \text{Tr}\left( G^{\mf}_k \Gamma_{\ts} \right)$ where
$G^{\mf}_k=(i\omega_n-H^0_\ka+\Delta^{\ts} \Gamma_{\ts})^{-1}$ is the MF
Green's function. Thus, the MF solution simply corrects the mass term $M$ so that
the model describes a continuous TQPT occurring at the critical line $M-\frac{1}{2}\Delta^{\ts}_{\mf}= 2$~\footnote{In this work we consider values of
  the interaction which do not lead to a change of sign in the
  renormalized mass term} as reported in the phase diagram
of \figu{fig1}(b). 
All along this line the energy gap closes through the formation of a gapless Dirac node at the $\Gamma$
point as in the non-interacting model.
As we show in \figu{fig1}(a) (dotted line), the orbital polarization smoothly evolves across the topological transition.  Coherently with the above scenario, a direct inspection of the MF free energy shows only one minimum for every value of $g_{\ts}$ and $M$.

In this work we overcome the limitations of the MF by approximating the exact free energy functional 
with a second-order expansion in the fluctuating fields, whose
coefficients are variationally determined~\cite{Hertz1974PRB}. We
underline that this approach includes effects of higher order beyond
the  conventional strategy of analyzing Gaussian fluctuations around
the MF solution~\cite{Klein2018PRB}.
Most importantly, the expectation values of the fields can change with respect to MF, i.e. $\bar{\vD} \neq \bar{\vD}_{\mf}$. Assuming $\vD_q\rightarrow \bar{\vD} + \vD_q$ we can write
\begin{equation}
    F[\vD]  \simeq F^{(2)}[\vD] 
     = F[\bar{\vD}] 
    + \frac{1}{2} \sum_{abq}  \Delta^a_q A^{ab}_q \Delta^b_{-q}  
\label{Fgaussian}
\end{equation}
where the variational principle $\FF \leq \FF^{(2)} +
\feval{F[\vD]  - F^{(2)}[\vD]} $,
with  $\FF^{(2)}\!=\!-\frac{1}{\beta}\ln{\int    \DD\vD \ e^{-\beta \mathcal{N} F^{(2)}[\vD]}}$ 
leads to the new stationary condition $\!\feval{ \partial_{\Delta^a_q} F[\vD] }=0$ and
$A^{ab}_q\!=\!\feval{ \partial_{\Delta^a_q}\partial_{\Delta^b_{-q}}F[\vD]}$.
The symbol $\feval{\cdot}$ indicates that the averages over the possible configurations are calculated using the second-order probability density of the fluctuating field. 

The coefficients $A^{ab}$ in the free energy expansion \eqref{Fgaussian}
depend on the averaged dressed Green's function $\feval{\GG_{kq}[\vD]}$, which can not be calculated exactly. 
In order to circumvent this problem we can introduce an
auxiliary potential $\Sigma_k$ implicitly determined
by the condition~\cite{Kakehashi2002PRB,Melnikov2011JOPCM}:
$\frac{\partial \FF^{(2)}}{\partial \Sigma_k}=0$, 
which in turn implies that  $\feval{\GG_{kq}[\vD]}$ coincides with an 
interacting Green's function in which $\Sigma_k$ plays the role of a self-energy
$G_k = \left[i \omega_n + \mu - H_0(\vec{k}) -
  \Sigma_k \right]^{-1}$.

The stationary condition and the expression for $A^{ab}_q$ become
\begin{equation}
    \frac{\bar{\Delta}^a}{g_a} =
    \frac{1}{\beta \mathcal{N}} \text{Tr}\left(G_k\Gamma_a \right)
    ;\,
    A^{ab}_q  = \frac{\delta_{ab} }{g_a} - \chi_{ab}(q).
    \label{ABexpr}
\end{equation}
The first expression contains the fluctuation-corrected Green's function, hence it leads to corrected values of the $\bar{\vD}$, while the second can be seen as an optimized version of the Random-phase Approximation (RPA), as we shall discuss in the following~\cite{Klein2018PRB}.
Indeed $\chi_{ab}(q) = - \tfrac{1}{\beta \mathcal{N}} \sum_k \text{Tr}
\left[G_k\Gamma_a G_{k+q} \Gamma_b\right]$ is the susceptibility
matrix in the space of the different channels which, in presence of odd hybridization between the orbitals, has diagonal structure  $\chi_{aa}\delta_{ab}$.
Moreover, the symmetries of the
interaction ensure that $\chi_{\ss \ss}=\chi_{\ns \ns}$ and
$\chi_{\rs \rs}=\chi_{\ts \ts}$. 
$\Sigma_k$ can be written explicitly up to second order as  
\begin{equation}
  \Sigma_k = \bar{V} + \sum_{qa} G_{k-q} \feval{|\Delta^a_q|^2} 
  \label{Sigma}
\end{equation}
where 

\begin{equation}
  \feval{|  \Delta^a_q|^2} = \tfrac{1}{\beta \mathcal{N}}
  \left[ \frac{1}{g^{-1}_a  - \chi_{aa}(q)} - g_a \right].
  \label{dVq}
\end{equation}

The  diagonal form in the channel index allows us to analyze the contribution of each fluctuating term of the interaction to the potential $\Sigma_k$~\cite{Gunnarsson2015PRL}.
We emphasize that, formally, the second term in \eqref{Sigma} plays the same role of the one-loop quantum correction of the  effective Coleman-Weinberg potential~\cite{Coleman1973PRD}.

The Eqs.(\ref{ABexpr}) and (\ref{Sigma}) provide a closed system of non-linear
equations for $\Sigma_k$ and the bosonic fields $\bar{\vD}$. 
Previous studies suggest that the interaction
effects on the TQPT are mainly local and that non-local fluctuation play a minor role~\cite{Crippa2021PRB}. 
Thus, in order to further simplify the treatment in the following we will
assume a local $\Sigma_k \simeq \Sigma(i \omega_n) $.
We solve this system iteratively using a linear mixing algorithm 
which typically converges in $10-20$ steps.
The BZ is discretized with a linear grid of $20\times 20$ points and
the Matsubara axis with $L=8192$ frequencies using an effective inverse temperature  $\beta=500$. 
The convolution in \equ{Sigma} is evaluated using a FFT algorithm. We discuss the results obtained for $J/U=1/8$ and $\lambda=0.3$.

In Fig.\ref{fig1}(a) we show the evolution of the orbital polarization, obtained 
from the self-consistent value of the bosonic
field $\bar{T}_z=\bar{\Delta}^{\ts}/g_{\ts}$. The behavior at the
transition point $g^*_{\ts}(M)$ changes qualitatively according to the value of the bare mass 
$M$.
For a value close to $M=2$, i.e. the non-interacting transition point, the
orbital polarization is continuous with respect to the increasing
interaction $g_{\ts}$. This corresponds to a smooth modification of the
BI into a non-trivial insulator through the formation of a gapless
state at the TQPT.
Starting from a farther point, the orbital polarization displays a 
critical behavior at the transition characterized by a divergent susceptibility
$\partial_M \bar{T}_z$.
Beyond this point, for any value of $M$, the orbital polarization is
characterized by a discontinuous evolution across the topological
transition. This is in stark contrast with the continuous behavior obtained in MF for the same value of the mass, see \figu{fig1}(a), and it agrees with previous results obtained via Dynamical Mean-Field Theory (DMFT)~\cite{Amaricci2015PRL}. The agreement with DMFT is indeed even quantitative, as shown by the data reported in \figu{fig1}(b). 

These results can be summarised in a phase diagram in the plane $g_{\ts}$-$M$ 
(see \figu{fig1}(b)) where we compare the MF and fluctuation-corrected results for the TQPT.
The two transition lines remain close for small
values of the interaction $g_{\ts}$. 
Accordingly, the  
free energy functional $F$ displays a single
minimum as a function of  $\Delta^{\ts}$ (inset A in \figu{fig1}(b)).  
However, upon increasing the interaction strength the two curves start
deviating significantly signalling a crucial impact of the fluctuations.  

A direct information about the contribution of the fluctuations in the different channels 
is reported in \figu{fig1}(c). The
fast increasing behavior with the interaction $g_{\ts}$ in all the channels  
stops at the topological transition towards the QSHI, where these quantities display a
discontinuous drop and a successive slow increase.    
The terms $\feval{|\Delta^a_q|^2}$ enter, through \eqref{dVq}, 
in $\Sigma(i \omega_n)$ giving it a dynamical nature which
significantly deviates from its static MF form $\bar{V}=-\bar{\Delta}^{\ts}\Gamma_{\ts}$.
This  results in a crucial shift of the self-consistent saddle point value of
the bosonic fields. 
Moreover, as discussed above, while the MF always describes a
continuous transition, in the corrected theory the  boundary line is continuous up
to a critical value $g^c_{\ts}$ of the interaction beyond which it becomes of first
order. This reflects in the behavior of the free energy near
the TQPT point in the  intermediate to strong coupling regime, i.e.
$g_{\ts}> g_{\ts}^c$.
In the insets B and C of \figu{fig1}(b) we compare the free energies of a QSHI state near the topological transition for, respectively, the
MF and the fluctuation-corrected approximation, where two minima are found:  A stable QSHI and a  metastable BI.

We thus find that a QCP separates the continuous from the discontinuous
regime on the topological transition line, where we also found a
divergent orbital susceptibility.  Indeed the uniform orbital susceptibility
$\X_{\ts}=\partial_M \bar{\Delta}^{\ts}_{q=0}$ reads, using \equ{ABexpr},
\begin{equation}
  \X_{\ts} = \frac{-2 g_{\ts} \chi_{{\ts} {\ts}} (0) }{1- g_{\ts}
   \chi_{{\ts} {\ts}} (0) + \lambda},
 \label{final}
\end{equation}
which reminds of the RPA result with a correction $\lambda$ that stems from the implicit dependence of $\Sigma(i \omega_n)$ on the orbital polarization. This quantity also  accounts for the contributions of all the other channels of the interaction through the
expression of $\Sigma(i \omega_n)$, see \equ{Sigma}~\cite{Fanfarillo2012PRB}.
At the MF level we find $\Sigma(i \omega_n) = -\bar{\Delta}^{\ts}\Gamma_{\ts}$, so that $\lambda=0$ and $\X_{\ts}$
reduces to the RPA form. For a Hubbard-Kanamori interaction the RPA $\X_{\ts}$ 
diverges only either for negative $U$ or negative $M$ (which lead to different physics), in agreement with the continuous TQPT we always find. 
Note that this result holds true also when all the coupling constants but $g_{\ts}$ vanish and the interaction reduces to $\frac{g_{\ts}}{2}\hat{T}_z^2$~\cite{Roy2016PRB}.

\begin{figure}
\includegraphics[width=\linewidth]{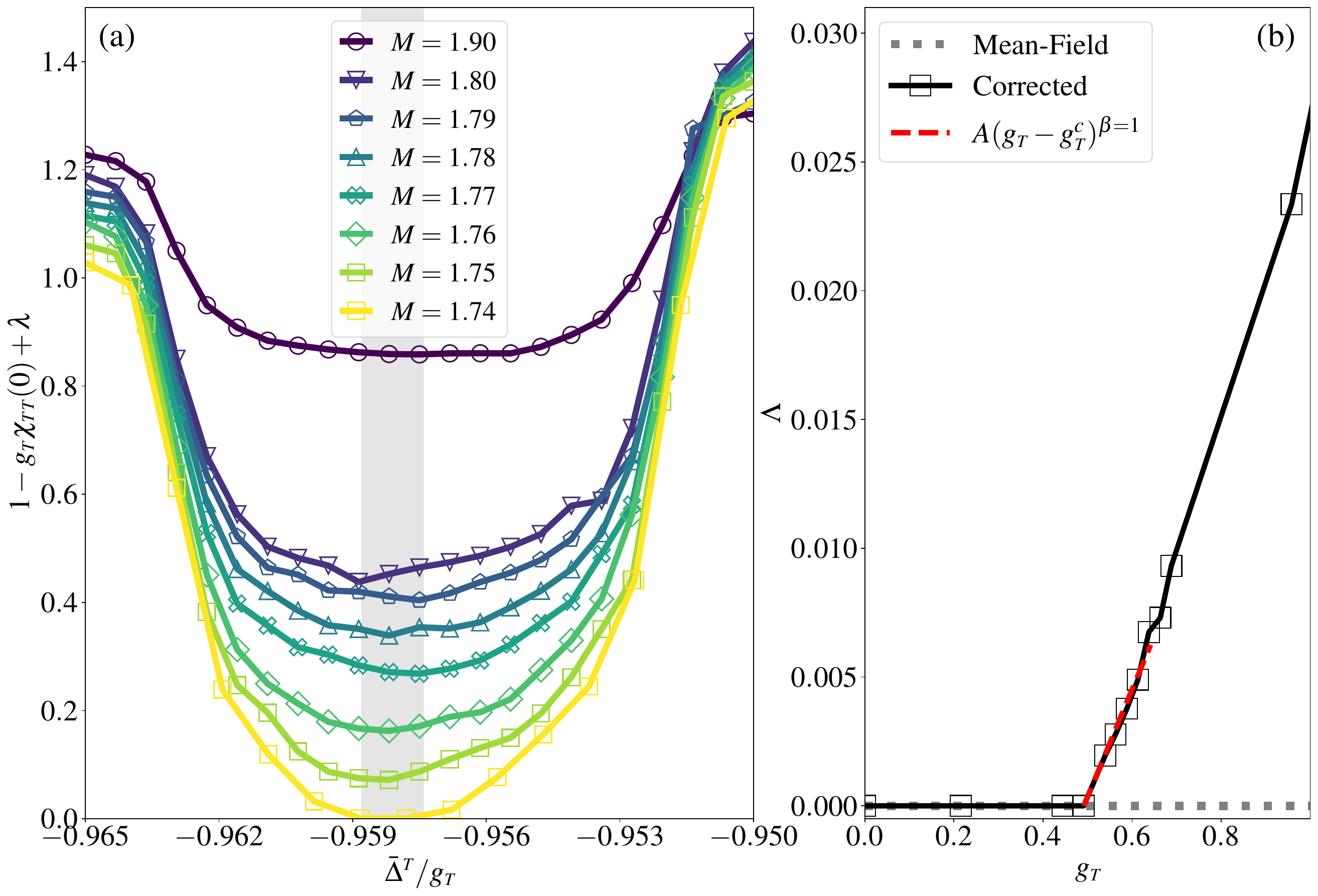}
\caption{(Color online)
  (a) The denominator of \equ{final} as a function of
  the  orbital polarization $\bar{\Delta}^{\ts}/g_{\ts}$ across the
  TQPT in the fluctuation-corrected approximation and for different
  values of the bare mass $M$.
  The narrow grey stripe indicates the minimum with its numerical uncertainty.
  (b) The gap $\Lambda$ along the TQPT
  line as a function of the interaction $g_{\ts}$. MF is the dotted grey
  line, while fluctuation-corrected results are indicated by open symbols and solid line. The (red) dashed line is a linear fit
  $A(g_{\ts}-g^c_{\ts})^{\beta=1}$ ($A\simeq0.042$) of the critical behavior.}
\label{fig2}
\end{figure}

In order to compute $\lambda$ we consider the
zero-frequency limit of $\Sigma(i \omega_n)$ where we obtain
$\lambda \simeq \chi_{{\ts}{\ts}} \partial_{\bar{T}_z} \Lambda $. Since  $\partial_{\bar{T}_z} \Lambda < 0$, this correction is negative and tends to enhance $\X_{\ts}$.
For a fixed value of $M$ the TQPT corresponds to the maximum of the response
function which connects the continuous transition to a Widom line~\cite{Xu2005PNAS,Simeoni2010NP,Sordi2012SR}. The minimum of the denominator in 
\equ{final} approaches zero when we reach the QCP, as shown in \figu{fig2}(a).
As we discussed above, in a non-interacting TQPT the spectral gap
closes at the transition.
We now compute the gap in our scheme from  the zero-frequency
limit of the self-energy $\Lambda=\Re\Sigma(i\omega_n\!\rightarrow\!0)-\bar{V}$.
In \figu{fig2}(b) we report the behavior of $\Lambda$ at the TQPT
as a function of the interaction strength $g_{\ts}$. While in MF the gap is always zero, including the fluctuations we find a finite gap above the QCP ($g_{\ts} > g^c_{\ts}$). 
A finite value of the  gap $\Lambda$ corresponds to give a mass to the
Dirac fermions at the boundary line.
This is consistent with a spontaneous symmetric mass generation
process~\cite{Slagle2015PRB,Wang2022S,You2018PRX}.
The presence of such finite gap (or mass) makes it
impossible to continuously connect the trivial with the non-trivial
phase and leads to a first-order TQPT.
In addition, we find numerically that the critical
behavior near the QCP falls in the Gross-Neveu universality
class~\cite{Slagle2015PRB,Liu2021PRB}, with an  estimated critical
exponent $\beta\simeq 1$.

In this work, using a non-perturbative analytical approach to include
interactions in the BHZ model, we have demonstrated
the crucial role of fluctuations in the different local particle-hole
channels to qualitatively change the nature of the TQPT.
Within a mean-field the interactions only lead to a renormalization of
the bare mass of the model (coupled to the orbital polarization) so
that the TQPT has the same character of the non-interacting
model. Within our approach the fluctuation contributions change the MF
parameters and lead to a discontinuous TQPT for large interactions
with a QCP separating the continuous and discontinuous branches.
This effect is intrinsically related to a
spontaneous gap opening (mass formation) for the otherwise gapless Dirac nodes at TQPT point without any symmetry breaking. The gap follows a 
Gross-Neveu critical behavior. This process of spontaneous mass
generation takes place through a condensed matter analog of the CW
mechanism in which one-loop quantum fluctuations lead to a mass
without symmetry breaking~\cite{Coleman1973PRD}.
We expect that our mechanism can be applied also to other models for topological phase transitions, but also to a wider class of phenomena.  A natural example is that of  Lifshitz transitions in 
interacting electronic systems, where the continuous deformation of the
Fermi surface topology characteristic of non-interacting systems is expected to share the same destiny of the TQPT, i.e., to become discontinuous for large interactions.

\paragraph*{Acknowledgments -- } 
F.P. is grateful to P.Coleman and G.Perosa for illuminating discussions specially
concerning Ref.~\cite{Coleman1973PRD}. We thank M. Fabrizio, G. Mazza, G. Sangiovanni, J.C. Budich, B. Trauzettel for useful discussions.  
We acknowledge financial support of MUR via PRIN 2017 (Prot.~20172H2SC4
005), PRIN 2020 (Prot.~2020JLZ52N 002) programs, PRIN 2022
(Prot. 20228YCYY7),  National Recovery and Resilience Plan (NRRP) MUR
Project No.~PE0000023-NQSTI and ICSC–Centro Nazionale di Ricerca in
High Performance Computing, Big Data and Quantum Computing, funded by
European Union – NextGenerationEU (Grant number CN00000013) -  Mission
4 Component 2 Investments 1.3 and 1.4.
\bibliography{references}

\end{document}